\documentclass[prd,showpacs,superscriptaddress,nofootinbib,floatfix,11pt]{revtex4-2}
\usepackage[utf8]{inputenc}
\usepackage{amsmath,amssymb}
\usepackage{slashed}
\usepackage{subfigure}
\usepackage[colorlinks=true,
breaklinks=true,
urlcolor=magenta,
citecolor=blue]{hyperref}
\usepackage[usenames,dvipsnames]{color}
\usepackage{appendix}
\usepackage{braket,bm}
\usepackage{multirow}
\usepackage{enumitem}
\usepackage{color}
\usepackage{slashed}
\usepackage{orcidlink}
\usepackage{graphicx}
\usepackage{tikz}
\usepackage{booktabs}
\usepackage{array}
\usepackage{overpic}
\usepackage{float}
\usepackage{threeparttable}
\usepackage{pifont}
\usepackage{bbding}
\usepackage{utfsym}
\allowdisplaybreaks[4]

\newcommand{\zjwl}{\affiliation{College of Information and Intelligence Engineering, Zhejiang Wanli University, Zhejiang 315101, China}}

\newcommand{\nbu}{\affiliation{Physics Department, Ningbo University, Zhejiang 315211, China}}

\newcommand{\kmu}{\affiliation{School of Physics Science and Technology, Kunming University, Kunming 650214, China}}

\begin{document}
\title{Possible Bound States in the $D^\ast\bar D^\ast$/$B^\ast\bar B^\ast$ and $D^\ast D^\ast$/$\bar B^\ast\bar B^\ast$ Systems within the Bethe-Salpeter Formalism}
\author{Ce Li}\email{2511690036@nbu.edu.cn}
\nbu

\author{Jing-Juan Qi\orcidlink{0000-0002-9260-9408}}\email{qijj@mail.bnu.edu.cn}
\zjwl\nbu

\author{Zhu-Feng Zhang\orcidlink{0000-0002-4418-6965}}\email{zhangzhufeng@nbu.edu.cn}
\nbu

\author{Zhen-Yang Wang\orcidlink{0000-0002-4074-7892}}\email{Corresponding authors:wangzhenyang@nbu.edu.cn}
\nbu

\author{Xin-Heng Guo\orcidlink{0000-0002-9309-9112}}\email{Corresponding authors:xhguo@bnu.edu.cn}
\kmu

\begin{abstract}
We investigate possible $S$-wave bound states in the $D^\ast\bar D^\ast$, $B^\ast\bar B^\ast$, $D^\ast D^\ast$, and $\bar B^\ast\bar B^\ast$ systems within the Bethe-Salpeter formalism using one-boson-exchange interactions. Bound state solutions are obtained in the isoscalar hidden-heavy systems with $J^{PC}=0^{++}$, $1^{+-}$, and $2^{++}$, whereas no isovector solutions are found within the parameter range considered. For the doubly heavy systems, solutions are obtained in the allowed $I(J^P)=0(1^+)$, $1(0^+)$, and $1(2^+)$ systems, although the $1(0^+)$ solution requires a comparatively large cutoff parameter. The bottom systems are bounded more favorably than their charmed counterparts because of their larger reduced masses.
\end{abstract}
\maketitle
\newpage

\section{Introduction}
\label{intro}
Quantum chromodynamics (QCD) is the fundamental theory of the strong interaction. Conventional low-lying mesons and baryons are well described as $q\bar q$ and $qqq$ states within the constituent quark model~\cite{Gell-Mann:1964ewy,Zweig:1964ruk}. Nevertheless, QCD also permits color-singlet configurations beyond the conventional quark-model picture, including multiquark states,
glueballs, and hybrid hadrons. Over the past two decades, numerous candidates for exotic hadrons have been observed experimentally~\cite{ParticleDataGroup:2024cfk}, but their internal structures have not yet been conclusively established. Among the proposed interpretations of exotic hadrons, compact multiquark states and hadronic molecules are two of the most widely discussed scenarios. Distinguishing between these configurations is important for understanding both exotic hadron spectroscopy and the nonperturbative dynamics of QCD.

The $X(3872)$, first observed by the Belle Collaboration in $B^\pm\to K^\pm\pi^+\pi^-J/\psi$ decays~\cite{Belle:2003nnu}, is one of the best known exotic hadron candidates. Its mass lies extremely close to the $D^0\bar D^{\ast0}$ threshold, which has motivated extensive studies of the $X(3872)$ as a $D\bar D^\ast$ hadronic molecule~\cite{Swanson:2003tb,Kalashnikova:2005ui,Zhang:2009vs,Chen:2009zzi,Ding:2009vj,Liu:2008qb,Lee:2009hy,Guo:2014taa,He:2014nya,Zhao:2014gqa,Guo:2013sya,Wang:2017dcq,Lu:2025zae,Li:2024pfg,Wong:2003xk}, a compact tetraquark~\cite{Maiani:2004vq,Ebert:2005nc,Carlucci:2007um,Vijande:2007fc,Maiani:2007vr,Dubnicka:2010kz,Wang:2013vex,Padmanath:2015era,Anwar:2018sol,Wallbott:2019dng}, or a mixture of molecular and charmonium components. Its proximity to an $S$-wave two hadron threshold makes the molecular interpretation particularly compelling. Several other exotic candidates also lie close to open heavy flavor hadron thresholds. Examples include the $Z_c(3900)$~\cite{BESIII:2013ris} near the $D\bar D^\ast$ threshold, the $Z_c(4020)$~\cite{BESIII:2013ouc}, $X(3960)$~\cite{Belle:2005lik}, and $X(4160)$~\cite{Belle:2007woe}  near the $D^\ast\bar D^\ast$ threshold, the $T_{cc}^+(3875)$~\cite{LHCb:2021vvq,LHCb:2021auc} near the $DD^\ast$ threshold, and the $Z_b(10610)$ and $Z_b(10650)$~\cite{Belle:2011aa} near the $B\bar B^\ast$ and $B^\ast\bar B^\ast$ thresholds, respectively. Such threshold proximity suggests that the interactions between the corresponding hadrons may play an important role in the formation of these states~\cite{Dong:2020hxe}. It has therefore motivated extensive investigations of their molecular structures, heavy quark spin partners, decay modes, and production mechanisms~\cite{Chen:2016qju,Guo:2017jvc,Brambilla:2019esw,Liu:2024uxn,Ding:2020dio,Dong:2021juy,Dong:2021bvy,Ke:2021rxd,Yalikun:2025ssz,Abreu:2025mhl,Cai:2025inq,Dias:2024zfh,Liu:2024ziu,Deng:2024pep,Wu:2023rrp,Wang:2022aiu}.

In particular, if the $X(3872)$ is predominantly a $D\bar D^\ast$ molecule, heavy quark spin symmetry (HQSS) suggests the possible existence of related $D^\ast\bar D^\ast$ molecular states~\cite{Nieves:2012tt,Hidalgo-Duque:2012rqv,Baru:2016iwj}. Doubly heavy $D^\ast D^\ast$ and $\bar B^\ast\bar B^\ast$ systems are also of considerable interest~\cite{Zhang:2025rlg,Ren:2026wzt}, especially following the observation of the $T_{cc}^+(3875)$. Compared with charmed systems, bottom systems have larger reduced masses and therefore bigger binding energy, which may make molecular binding tighter.

In this work, we systematically investigate possible $S$-wave bound states in the $D^\ast\bar D^\ast$, $B^\ast\bar B^\ast$, $D^\ast D^\ast$, and $\bar B^\ast\bar B^\ast$ systems within the Bethe-Salpeter (BS) formalism. The BS equations are solved in the ladder and instantaneous approximations. The interaction kernels are constructed from one-boson exchanges including the $\sigma$, pseudoscalar ($\pi$ and $\eta$), and vector ($\rho$ and $\omega$) mesons. This approach is similar to that previously applied to systems consisting of a heavy vector meson and a heavy pseudoscalar meson~\cite{Zhao:2021cvg}, as well as to heavy baryonium and heavy dibaryon systems~\cite{Qi:2024dqz}.

The remainder of this paper is organized as follows. In Sec.~\ref{Formalism}, we introduce the effective Lagrangians, interaction kernels, and BS equations for the hidden-heavy and doubly heavy systems. The numerical results and discussions are presented in Sec.~\ref{res and dis}. Finally, Sec.~\ref{conclusion} summarizes our main conclusions.

\section{Formalism}
\label{Formalism}
For the light quark isospin conventions, we adopt the following ones:
\[\vert u\rangle=\left\vert \frac12,\frac12\right\rangle,\quad\vert d\rangle=\left\vert \frac12,-\frac12\right\rangle,\] 
and for the corresponding antiquarks:
\[ \vert \bar{u}\rangle=\left\vert \frac12,-\frac12\right\rangle,\quad \vert \bar{d}\rangle=-\left\vert \frac12,\frac12\right\rangle.\]
Accordingly, the charmed vector mesons form the isospin doublets
\[
D^\ast=(-D^{\ast+},D^{\ast0})^T,\quad
\bar D^\ast=(\bar D^{\ast0},D^{\ast-})^T.
\] 
The relative phases in these definitions are conventional but must
be used consistently in the flavor wave functions, interaction
vertices, and isospin factors. The corresponding field operators are 
\begin{equation}
    \begin{split}
        D^{\ast\mu}&=\int\frac{d^3k}{(2\pi)^3}\frac{1}{\sqrt{2E_{\mathbf{k}}}}\sum_\lambda\left[a(\mathbf{k},\lambda)\epsilon^\mu(\mathbf{k},\lambda)e^{-ik\cdot x}+a^\dag(\mathbf{k},\lambda)\epsilon^{\mu\ast}(\mathbf{k},\lambda)e^{ik\cdot x}\right],\\
        \bar{D}^{\ast\mu}&=\int\frac{d^3k}{(2\pi)^3}\frac{1}{\sqrt{2E_{\mathbf{k}}}}\sum_\lambda\left[b(\mathbf{k},\lambda)\epsilon^\mu(\mathbf{k},\lambda)e^{-ik\cdot x}+b^\dag(\mathbf{k},\lambda)\epsilon^{\mu\ast}(\mathbf{k},\lambda)e^{ik\cdot x}\right].\\
    \end{split}
\end{equation}

Using the above phase conventions and the corresponding Clebsch-Gordan coefficients, the flavor wave functions of the $D^\ast\bar D^\ast$ and $D^\ast D^\ast$ systems are constructed as
\begin{equation}
    \begin{split}
        |T_{c\bar{c}};{I=0,I_3=0}\rangle&=-\frac{1}{\sqrt{2}}\left(|D^{\ast+}D^{\ast-}\rangle+|D^{\ast0}\bar{D}^{\ast0}\rangle\right),\\
        |T_{c\bar{c}};{I=1,I_3=1}\rangle&=-|D^{\ast+}\bar{D}^{\ast0}\rangle,\\
        |T_{c\bar{c}};{I=1,I_3=0}\rangle&=-\frac{1}{\sqrt{2}}\left(|D^{\ast+}D^{\ast-}\rangle-|D^{\ast0}\bar{D}^{\ast0}\rangle\right),\\
        |T_{c\bar{c}};{I=1,I_3=-1}\rangle&=|D^{\ast0}D^{\ast-}\rangle,\\
    \end{split}
\end{equation}
and
\begin{equation}
    \begin{split}
        |T_{c{c}};{I=0,I_3=0}\rangle&=-\frac{1}{\sqrt{2}}\left(|D^{\ast+}D^{\ast0}\rangle-|D^{\ast0}{D}^{\ast+}\rangle\right),\\
        |T_{c{c}};{I=1,I_3=1}\rangle&=|D^{\ast+}{D}^{\ast+}\rangle,\\
        |T_{c{c}};{I=1,I_3=0}\rangle&=-\frac{1}{\sqrt{2}}\left(|D^{\ast+}D^{\ast0}\rangle+|D^{\ast0}{D}^{\ast+}\rangle\right),\\
        |T_{c{c}};{I=1,I_3=-1}\rangle&=|D^{\ast0}D^{\ast0}\rangle.\\
    \end{split}
\end{equation}

For an $S$-wave $D^\ast\bar{D}^\ast$ system, the orbital angular momentum is $L=0$ and the total parity is positive. The charge conjugation parity is $C=(-1)^{L+S}=(-1)^S$. The allowed quantum numbers are therefore $J^{PC}=0^{++}$, $1^{+-}$, $2^{++}$. Because the meson and antimeson are distinguishable, Bose symmetry does not impose an additional restriction. Both isoscalar and isovector configurations are thus allowed:
\[
I=0,1:\quad J^{PC}=0^{++},\ 1^{+-},\ 2^{++}.
\]

For an $S$-wave $D^\ast D^\ast$ system, charge conjugation is not a
good quantum number. Since the two vector mesons are identical bosons in the isospin-symmetric limit, the total wave function must be symmetric under particle interchange. The $L=0$ spatial wave function is symmetric, whereas the spin wave function is symmetric for $S=0,2$ and antisymmetric for $S=1$. Combining the spin and isospin wave functions gives the following allowed systems:
\[
I(J^P)=0(1^+),\quad 1(0^+),\quad 1(2^+).
\]
The same restrictions apply to the corresponding doubly bottom system.

The interactions of charmed vector mesons with light scalar, pseudoscalar, and vector mesons are described using effective Lagrangians consistent with heavy quark symmetry and chiral symmetry. We include the exchanges of the scalar meson $\sigma$, the pseudoscalar mesons $\pi$ and $\eta$, and the vector mesons $\rho$ and $\omega$. The relevant interaction Lagrangians are \cite{Ding:2009zq}
\begin{equation}\label{lagrangian}
    \begin{split}
        \mathcal{L}_{D^\ast D^\ast\sigma}&=g_{D^\ast D^\ast\sigma}D_a^{\ast\mu}D_{a\mu}^{\ast\dag}\sigma,+g_{\bar{D}^\ast \bar{D}^\ast\sigma}\bar{D}_a^{\ast\mu}\bar{D}_{a\mu}^{\ast\dag}\sigma,\\
        \mathcal{L}_{D^\ast D^\ast\mathcal{P}}&=g_{D^\ast D^\ast\mathcal{P}}\epsilon_{\mu\nu\alpha\beta}D_a^{\ast\nu\dag}\overleftrightarrow{\partial}^\beta D_b^{\ast\mu}\partial^\alpha\mathcal{P}_{ab},+g_{\bar{D}^\ast \bar{D}^\ast\mathcal{P}}\epsilon_{\mu\nu\alpha\beta}\bar{D}_a^{\ast\mu\dag}\overleftrightarrow{\partial}^\beta \bar{D}_b^{\ast\nu}\partial^\alpha\mathcal{P}_{ab},\\       
       \mathcal{L}_{D^\ast D^\ast\mathcal{V}}&=ig_{D^\ast D^\ast\mathcal{V}}\left(D^\ast_{b\nu}\overleftrightarrow{\partial}_\mu D^{\ast\nu\dag}_a\right)\mathcal{V}_{ba}^\mu+ig'_{D^\ast D^\ast\mathcal{V}}\left(D_b^{\ast\mu}D_a^{\ast\nu\dag}-D_a^{\ast\mu\dag}D_b^{\ast\nu}\right)(\partial_\mu\mathcal{V}_\nu-\partial_\nu\mathcal{V}_\mu)_{ba},\\
       &+ig_{\bar{D}^\ast \bar{D}^\ast\mathcal{V}}\left(\bar{D}^\ast_{b\nu}\overleftrightarrow{\partial}_\mu \bar{D}^{\ast\nu\dag}_a\right)\mathcal{V}_{ba}^\mu+ig'_{\bar{D}^\ast \bar{D}^\ast\mathcal{V}}\left(\bar{D}_b^{\ast\mu}\bar{D}_a^{\ast\nu\dag}-\bar{D}_a^{\ast\mu\dag}\bar{D}_b^{\ast\nu}\right)(\partial_\mu\mathcal{V}_\nu-\partial_\nu\mathcal{V}_\mu)_{ba},
    \end{split}
\end{equation}
where $A\overleftrightarrow{\partial}_\mu B=A(\partial_\mu B)-(\partial_\mu A)B$, $\mathcal{P}$ and $\mathcal{V}$ are $3\times3$ matrices of the pseudoscalar octet and vector meson nonet,
\begin{equation}
\label{pseudoscalar}
\mathcal{P}=\left(
\begin{array}{ccc} \frac{\pi^0}{\sqrt{2}}+\frac{\eta}{\sqrt{6}} &\pi^+&K^+\\
                  \pi^-& -\frac{\pi^0}{\sqrt{2}}+\frac{\eta}{\sqrt{6}} &K^0\\
                     K^- &  \bar{K}^0 &    -\sqrt{\frac{2}{3}}\eta  \\
\end{array} \right),
\end{equation}
and
\begin{equation}
\label{vector}
\mathcal{V}=\left(
\begin{array}{ccc} \frac{\omega}{\sqrt{2}}+\frac{\rho^0}{\sqrt{2}} &\rho^+&K^{\ast+}\\
                  \rho^-&\frac{\omega}{\sqrt{2}}-\frac{\rho^0}{\sqrt{2}}&K^{\ast0}\\
                     K^{\ast-} &  \bar{K}^{\ast0} &    \phi  \\
\end{array} \right).
\end{equation}

The coupling constants are as follows
\begin{equation}\nonumber
   \begin{split}
         g_{D^\ast D^\ast \sigma}&=-g_{\bar{D}^\ast \bar{D}^\ast \sigma}=2g_\sigma m_{D^\ast}\\
        g_{D^\ast D^\ast \mathcal{P}}&=-g_{\bar{D}^\ast \bar{D}^\ast \mathcal{P}}=\frac{g_{\pi}}{f_\pi},\\
        g_{D^\ast D^\ast \mathcal{V}}&=-g_{\bar{D}^\ast \bar{D}^\ast\mathcal{V}}=-\frac{\beta g_V}{\sqrt{2}},\\
        g'_{D^\ast D^\ast \mathcal{V}}&=-g'_{\bar{D}^\ast \bar{D}^\ast\mathcal{V}}=-\sqrt{2}\lambda g_Vm_{D^\ast}
    \end{split}
\end{equation}
and $g_\sigma=-\frac{g_{\pi}}{2\sqrt{6}}$ with $g_\pi=3.73$, $g_V\simeq5.8$ by  imposing the KSRF relations, $\beta$ is estimated to be about 0.9 by vector meson dominance, $\lambda=0.56 \text{GeV}^{-1}$, and $f_\pi=132$ MeV.

The wave functions and Lagrangians of the hidden-bottom and doubly bottom systems can be obtained analogously.

\subsection{$D^\ast\bar{D}^\ast$}
The BS wave function of the $D^\ast\bar D^\ast$ system is defined as
\begin{equation}
\begin{split}\label{BS wf hiddencharm}
\chi_P^{\mu\nu}(x_1,x_2,P)&=\langle0|TD^{\ast\mu}(x_1)\bar{D}^{\ast\nu}(x_2)|P\rangle,\\
&=e^{-iP X}\int\frac{d^4p}{(2\pi)^4}e^{-ip x}\chi_P^{\mu\nu}(p),\\
\end{split}
\end{equation}
where $P=Mv$ is the total momentum of the bound state, and $v$ is its velocity. $D^{\ast\mu}(x_1)$ and $\bar{D}^{\ast\nu}(x_2)$ are the field operators of the constituent particles at space coordinates $x_1$ and $x_2$, respectively. $X\equiv\lambda_1 x_1+\lambda_2x_2$ is the center of mass coordinate and $x\equiv x_1-x_2$ is the relative coordinate, with $\lambda_{1(2)}=m_{1(2)}/\left(m_1+m_2\right)$, where $m_{1(2)}$ is the mass of the constituent particle. Accordingly, the constituent momenta are $p_1=\lambda_1P+p$ and $p_2=\lambda_2P-p$.

The equation for the BS wave function can be derived from a four-point Green function which is express in terms of the  four-point truncated irreducible kernel $\bar{K}$,
\begin{equation}\label{BS hidden system}
    \chi_P^{\mu\nu}(p)=S^{\mu\alpha}(p_1)\int\frac{d^4q}{(2\pi)^4}\bar{K}_{\alpha\beta\lambda\tau}(P,p,q)\chi_P^{\lambda\tau}(q)S^{\nu\beta}(p_2),
\end{equation}
where $S^{\mu\alpha}(p_1)$ and $S^{\nu\beta}(p_2)$ are the propagators for the constituent particles. 

For convenience, the relative momentum is decomposed into longitudinal and transverse components with respect to the bound state velocity $v^\mu$: $p_l=p\cdot v$, $p_t^\mu=p^\mu-(p\cdot v)v^\mu$. Then the propagators can be express as 
\begin{equation}\label{propagator1}
        S^{\mu\alpha}(p_1)=-i\frac{g^{\mu\nu}-(\lambda_1Mv+p_lv+p_t)^\mu(\lambda_1Mv+p_lv+p_t)^\alpha/m_1^2}{(\lambda_1M+p_l)^2-w_1^2+i\epsilon},
\end{equation}
and
\begin{equation}\label{propagator2}
        S^{\nu\beta}(p_2)=-i\frac{g^{\nu\beta}-{(\lambda_2Mv-p_lv-p_t)^\nu (\lambda_2Mv-p_lv-p_t)^\beta}/{m_2^2}}{(\lambda_2M-p_l)^2-w_2^2+i\epsilon}.
\end{equation}

Because the centrifugal barrier generally disfavors higher partial waves near threshold~\cite{Guo:2017jvc}, we restrict the present analysis to $S$-wave configurations. The covariant structures of the BS wave functions
for $J^{PC}=0^{++}$, $1^{+-}$, and $2^{++}$ are written as
\begin{equation}\label{wf hidden system}
\begin{split}
    \chi^{\mu\nu}_{0^{++}}(P,p)&=\phi(p_l,p_t^2)g^{\mu\nu}_t,\\
    \chi^{\mu\nu}_{1^{+-}}(P,p)&=\psi(p_l,p_t^2)\epsilon^{\mu\nu\alpha\beta}v_\alpha\epsilon_\beta,\\
    \chi^{\mu\nu}_{2^{++}}(P,p)&=\eta(p_l,p_t^2)\xi^{\mu\nu},
\end{split}
\end{equation}
where $g_t^{\mu\nu}=g^{\mu\nu}-v^\mu v^\nu$. The polarization vector $\epsilon^\mu$ and polarization tensor $\xi^{\mu\nu}$ satisfy
\begin{equation}\label{polarization}
    \begin{split}
        P_\mu\epsilon^\mu&=0,\quad T^{\mu\nu}\equiv
        \sum_\epsilon\epsilon^\mu\epsilon^\nu=\frac{P^\mu P^\nu}{M^2}-g^{\mu\nu},\\
        \xi^{\mu\nu}&=\xi^{\nu\mu},\quad
        \xi^{\mu\nu}g_{\mu\nu}=0,\quad
        P_\mu\xi^{\mu\nu}=0,\quad
        \sum_\xi\xi^{\mu\nu}\xi^{\alpha\beta}=\frac{1}{2}\left(T^{\mu\alpha}T^{\nu\beta}+T^{\mu\beta}T^{\nu\alpha}\right)-\frac{1}{3}T^{\mu\nu}T^{\alpha\beta}.
    \end{split}
\end{equation}

Based on the effective Lagrangians in Eq.\eqref{lagrangian}, the tree-level $t$-channel interaction kernels in the ladder approximation are given by
\begin{equation}\label{kernel hidden system}
    \begin{split}
        \bar{K}_\sigma^{\alpha\beta\lambda\tau}=&-C_\sigma^I g_{D^\ast D^\ast\sigma}g_{\bar{D}^\ast \bar{D}^\ast\sigma}g^{\alpha\lambda}\Delta(k,m_\sigma)g^{\beta\tau},\\ 
        \bar{K}_\mathcal{P}^{\alpha\beta\lambda\tau}=&-C_\mathcal{P}^I g_{D^\ast D^\ast\mathcal{P}}^2\epsilon_{\alpha\lambda\kappa\delta}\left(p_1+q_1\right)^\delta k^\kappa \Delta(k,m_{{\mathcal{P}}})\epsilon_{\tau\beta\gamma\zeta}\left(p_2+q_2\right)^\zeta k^\gamma,\\
    \bar{K}_\mathcal{V}^{\alpha\beta\lambda\tau}=&C_\mathcal{V}^I\left[g_{D^\ast D^\ast\mathcal{V}}(p_1+q_1)_\sigma g_{\alpha\lambda}+2g'_{D^\ast D^\ast\mathcal{V}}\left(k_\alpha g_{\sigma\lambda}-k_\lambda g_{\sigma\alpha}\right)\right]\Delta^{\sigma\delta}(k,m_{\mathcal{V}})\\
        &\times\left[g_{{D}^\ast {D}^\ast\mathcal{V}}(p_2+q_2)_\delta g_{\tau\beta}+2g'_{{D}^\ast {D}^\ast\mathcal{V}}(k_\tau g_{\delta\beta}-k_\beta g_{\delta\tau})\right].
    \end{split}
\end{equation}
Here, $\Delta(k,m_\sigma)$, $\Delta(k,m_{{\mathcal{P}}})$, and $\Delta^{\sigma\delta}(k,m_{\mathcal{V}})$ denote the propagators of the exchanged scalar, pseudoscalar, and vector mesons, respectively, $k=p-q$ is the transferred momentum. The coefficients $C^I_\sigma$, $C^I_P$, and $C^I_V$ are the corresponding isospin factors listed in Table~\ref{Isospin-factor}. Because only isospin-conserving strong interactions are considered, the BS equation depends on the total isospin $I$ but not on its third component $I_3$.

\subsection{$D^\ast D^\ast$}
For the doubly charmed system composed of double vector mesons, the general form of the BS equation has similar form to that of the hidden-charmed system, which in the momentum space is 
\begin{equation}\label{BS open system}
    \chi^{\mu\nu}_P(p)=S^{\mu\alpha}(p_1)\int\frac{d^4q}{(2\pi)^4}\bar{K}_{\alpha\beta\lambda\tau}(P,p,q)\chi^{\lambda\tau}_P(q)S^{\nu\beta}(p_2),
\end{equation}
with the BS wave function being defined as 
\begin{equation}
    \chi_P^{\mu\nu}(x_1,x_2)=\bra{0}TD^{\ast\mu}(x_1)D^{\ast\nu}(x_2)\ket{P}.
\end{equation}

The propagators $S^{\mu\alpha}(p_1)$ and $S^{\nu\beta}(p_2)$ for the two constituent particles $D^\ast$ have the same forms as those in Eqs. \eqref{propagator1} and \eqref{propagator2}, respectively.

The $t$-channel interaction kernels of the $D^\ast D^\ast$ system at the tree level are also obtained from the Lagrangians \eqref{lagrangian} as
\begin{equation}\label{kernel open system}
    \begin{split}
        \bar{K}_\sigma^{\alpha\beta\lambda\tau}(P,p,q)=&-C_\sigma^I g_{D^\ast D^\ast\sigma}g^{\alpha\lambda}\Delta(k,m_\sigma)g^{\beta\tau},\\
        \bar{K}_\mathcal{P}^{\alpha\beta\lambda\tau}(P,p,q)=&C_\mathcal{P}^Ig_{D^\ast D^\ast\mathcal{P}}^2\epsilon^{\alpha\lambda\kappa\delta}\left(p_1+q_1\right)_\delta k_\kappa\Delta(k,m_{\mathcal{P}})\epsilon^{\beta\tau\gamma\zeta}\left(p_2+q_2\right)_\zeta k_\gamma,\\
        \bar{K}_\mathcal{V}^{\alpha\beta\lambda\tau}(P,p,q)=&-C_\mathcal{V}^I\left[g_{D^\ast D^\ast\mathcal{V}}\left(p_1+q_1\right)^\sigma g^{\alpha\lambda}+2g'_{D^\ast D^\ast\mathcal{V}}\left(k^\alpha g^{\lambda\sigma}-k^\lambda g^{\alpha\sigma}\right)\right]\Delta_{\sigma\delta}(k,m_\mathcal{V})\\
        &\times\left[g_{D^\ast D^\ast\mathcal{V}}\left(p_2+q_2\right)^\delta g^{\beta\tau}+2g'_{D^\ast D^\ast\mathcal{V}}\left(-k^\beta g^{\delta\tau}+k^\tau g^{\beta\delta}\right)\right].
    \end{split}
\end{equation}
The isospin coefficients $C^I_\sigma$, $C^I_P$, and $C^I_V$ are also given in Table~\ref{Isospin-factor}.

Considering the $D^\ast\bar{D}^\ast$ and $D^\ast D^\ast$ are both composed of two charmed vector mesons, their Lorentz tensor structure of the $S$-wave covariant wave function is exactly the same,
\begin{equation}\label{wf open system}
\begin{split}
    \chi^{\mu\nu}_{0^{+}}(P,p)&=\phi(p_l,p_t^2)g^{\mu\nu}_t,\\
     \chi^{\mu\nu}_{1^{+}}(P,p)&=\psi(p_l,p_t^2)\epsilon^{\mu\nu\alpha\beta}v_\alpha\epsilon_\beta,\\
     \chi^{\mu\nu}_{2^{+}}(P,p)&=\eta(p_l,p_t^2)\xi^{\mu\nu}.
\end{split}
\end{equation}

\begin{table}[h]
\renewcommand{\arraystretch}{1.3}
\centering
\caption{Isospin factors.}\label{Isospin-factor}
\begin{tabular*}{\textwidth}{@{\extracolsep{\fill}}lcccccc}
\hline
\hline 
     &   &  $C_\sigma$ &   $C_\pi$   &    $C_\eta$  &   $C_\rho$   &    $C_\omega$  \\
\hline
              & & $\sigma$  & $\pi$ & $\eta$  & $\rho$ & $\omega$\\
\multirow{2}{*}{$D^\ast\bar{D}^\ast$}&$I=0$  &    1   &   3/2  &   1/6        &   3/2   &     1/2    \\

&$I=1$   &    1    &   -1/2   &   1/6       &   -1/2  &     1/2  \\

\multirow{2}{*}{$D^\ast D^\ast$}&$I=0$    &    1   &   -3/2   &   1/6       &   -3/2  &     1/2  \\

&$I=1$   &    1  &   1/2   &   1/6       &   1/2  &     1/2  \\
\hline
\hline
\end{tabular*}
\end{table}

To account phenomenologically for the finite size of the constituent hadrons and to regularize the high-momentum behavior of the interaction kernel, a monopole form factor is introduced at each interaction vertex:
\begin{equation}
    F(k^2)=\frac{\Lambda^2-m^2}{\Lambda^2-k^2},
\end{equation}
where $m$ is the mass of the exchanged meson and $\Lambda$ is the corresponding cutoff. Since the exchanged mesons with different masses probe different interaction ranges, we parameterize the cutoff as $\Lambda=m+\alpha\Lambda_{\rm QCD}$ with $\Lambda_{\rm QCD}=220~{\rm MeV}$. The dimensionless parameter $\alpha$ encodes model-dependent short-distance and finite-size effects. It cannot be determined from
first principles within the present framework and is therefore treated as a phenomenological parameter.

\section{results and discussions}
\label{res and dis}
The meson masses used in the numerical calculations are taken from the Particle Data Group \cite{ParticleDataGroup:2024cfk} and are listed in Table~\ref{masses}. The only adjustable parameter in the present model is $\alpha$, which determines the cutoff associated with each exchanged meson through 
\[\Lambda=m+\alpha\Lambda_{\rm{QCD}},\quad\Lambda_{\mathrm{QCD}}=220\ \mathrm{MeV}.\]
Although $\alpha$ is expected to be of order unity on naturalness grounds, we vary it over the range $0.5\leq\alpha\leq10$ to examine the sensitivity of the bound-state solutions to the regulator.

We focus on shallow bound states with binding energies $E_b= m_1+m_2-M$ in the range $1\leq E_b\leq50$ MeV, where $m_1$ and $m_2$ are the constituent masses and $M$ is the mass of the bound state. After substituting the propagators, interaction kernels, and covariant wave functions into the BS equations, the longitudinal momentum $p_l$ integration is performed by contour integration. After carrying out the azimuthal integration analytically, the four-dimensional BS equation is reduced to a one-dimensional integral equation. The resulting equation is discretized using Gaussian quadrature and solved as a matrix eigenvalue equation. A bound-state solution is identified when the relevant eigenvalue reaches unity.

\begin{table}[h]
\renewcommand{\arraystretch}{1.3}
\centering
\caption{Masses (in MeV) of mesons.}
\begin{tabular*}{\textwidth}{@{\extracolsep{\fill}}ccccc}
\hline
\hline 
$m_\sigma$  &  $m_{\pi^\pm}$ & $m_{\pi^0}$ & $m_\rho$  & $m_\omega$    \\
 \hline
  500   & 139.57 & 134.977     &  775.26    & 782.66   \\ 
  \hline\hline
  $m_{D^{\ast\pm}}$  &  $m_{D^{\ast0}}$ & $m_{B^{\ast}}$ &   \\
  \hline
  2010.27 &2006.86 & 5324.75 &    \\
\hline\hline
\end{tabular*}\label{masses}
\end{table}

\subsection{$D^\ast \bar{D}^\ast$ and $B^\ast\bar{B}^\ast$ systems}

\begin{figure}[htb]
\centering
\subfigure[]{
\includegraphics[width=5cm]{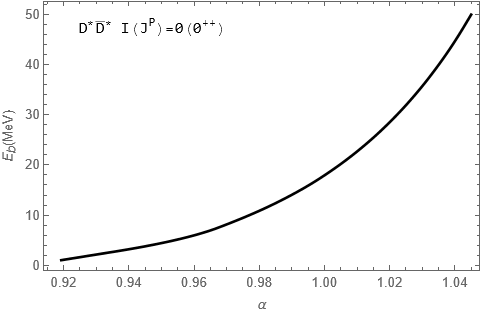}
}
\
\subfigure[]{
\includegraphics[width=5cm]{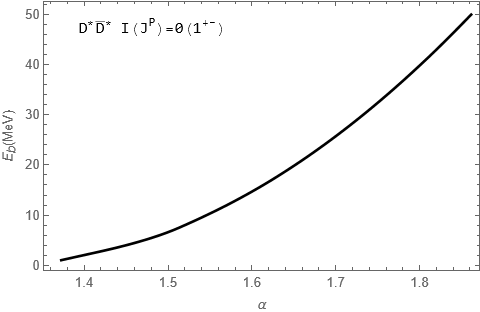}
}
\
\subfigure[]{
\includegraphics[width=5cm]{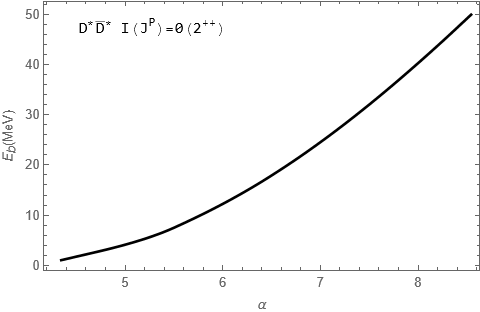}
}
\caption{Binding energy $E_b$ as a function of the cutoff parameter $\alpha$ for the possible $D^\ast\bar D^\ast$ bound states.}
\label{Hidden charmed systems}
\end{figure}

\begin{figure}[htb]
\centering
\subfigure[]{
\includegraphics[width=5cm]{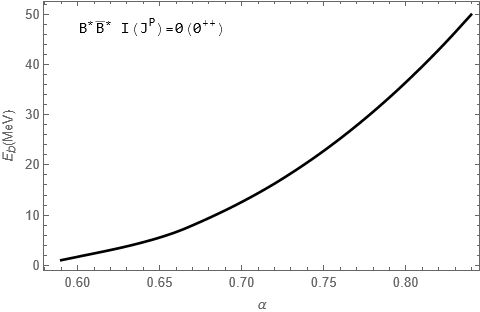}
}
\,
\subfigure[]{
\includegraphics[width=5cm]{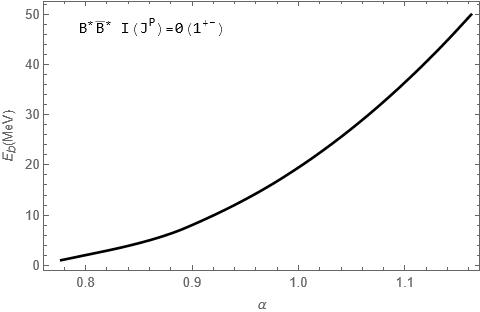}
}
\,
\subfigure[]{
\includegraphics[width=5cm]{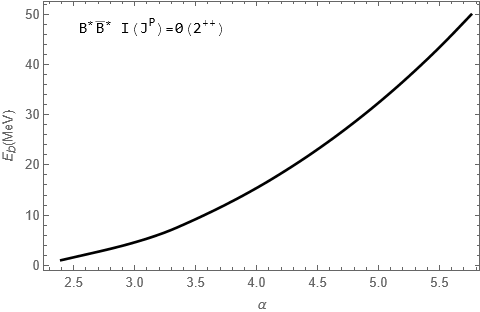}
}
\caption{Binding energy $E_b$ as a function of the cutoff parameter $\alpha$ for the possible $B^\ast\bar B^\ast$ bound states.}
\label{Hidden bottom systems}
\end{figure}

The dependence of the binding energies $E_b$ on the cutoff parameter $\alpha$ for the $D^\ast\bar D^\ast$ and $B^\ast\bar B^\ast$ systems are shown in Figs.~\ref{Hidden charmed systems} and~\ref{Hidden bottom systems}, respectively. Within the parameter range considered, bound state solutions are obtained in all the three isoscalar systems,
\[
I(J^{PC})=0(0^{++}),\quad 0(1^{+-}),\quad 0(2^{++}),
\]
provided that the cutoff parameter is sufficiently large. In contrast, no isovector bound-state solution is found in the present model.

For each bound system, the binding energy increases monotonically with $\alpha$. A larger $\alpha$ corresponds to a larger cutoff $\Lambda$, which reduces the suppression of high-momentum components by the form factor and thereby enhances the short-distance part of the regulated interaction. The resulting stronger attraction leads to a larger binding energy. However, solutions that require unusually large values of $\alpha$ are strongly regulator dependent and should be interpreted with caution, as they may reflect short-distance dynamics not explicitly included in the present one-boson-exchange kernel. For a given binding energy, the $B^\ast\bar B^\ast$ systems require smaller values of $\alpha$ than their $D^\ast\bar D^\ast$ counterparts. This behavior is mainly due to the larger reduced mass
of the bottom system, which lowers the kinetic energy contribution
and makes binding more favorable for the same attractive interaction.

Since the discovery of the $X(3872)$, its interpretation as a $D\bar{D}^\ast$ molecule has motivated extensive studies of its HQSS partners. Within the effective field theory framework constrained by the molecular interpretations of the $X(3872)$, $X(3915)$, and $Z_b(10610)$, several possible $D^\ast \bar{D}^\ast$ and $B^\ast\bar{B}^\ast$ bound states have been predicted. Assuming the $X(3872)$ and $X(3915)$ to be isoscalar molecules, Ref.~\cite{Nieves:2012tt} predicted $D^\ast\bar{D}^\ast$ bound states with $I(J^{PC})=0(1^{+-})$ and $0(2^{++})$, with masses of $3957\pm17$ MeV and $4012\pm3$ MeV, respectively. Based on the molecular interpretations of the $X(3872)$ and $Z_b(10610)$, Ref.~\cite{Guo:2013sya} predicted a $D^\ast\bar{D}^\ast$ bound state with $I(J^{PC})=0(2^{++})$ and a virtual state with $I(J^{PC})=1(1^{+-})$, with masses of $4012^{+4}_{-5}$ MeV and $4013^{+4}_{-11}$ MeV, respectively. Their $B^\ast\bar{B}^\ast$ counterparts were both predicted to be bound states, with masses of $10626^{+6}_{-7}$ MeV and $10648.1\pm2.1$ MeV, respectively. The same framework also predicted isovector $D^\ast\bar{D}^\ast$ molecular states with $I(J^{PC})=1(0^{++})$ and $1(1^{+-})$, with masses of $3953^{+24}_{-26}$ MeV and $3988^{+15}_{-17}$ MeV, respectively~\cite{Hidalgo-Duque:2012rqv}. All these results were obtained using a cutoff of $\Lambda=0.5~\mathrm{GeV}$. 

Other phenomenological approaches have also been applied to these systems. In the meson exchange model of Ref.~\cite{Liu:2009qhy}, isoscalar $D^\ast\bar{D}^\ast$ bound states with $J^{PC}=0^{++}, 1^{+-}, 2^{++}$ were obtained for $\Lambda\approx0.5$ GeV. By contrast, the isovector $0^{++}$ and $1^{+-}$ states requires substantially larger cutoffs, while no isovector $2^+$ bound state was found. The $B^\ast\bar{B}^\ast$ system exhibits the same qualitative pattern but requires smaller cutoffs for a given binding energy because of its larger reduced mass. In the chiral quark model analysis of Ref.~\cite{Liu:2008mi}, an isoscalar $D^\ast\bar{D}^\ast$ bound state was found only in the $J=2$ system, whereas isoscalar $B^\ast\bar{B}^\ast$ bound states were obtained in the $J=1$ and $2$ systems. Additional bound state systems were favored when the extended chiral quark model was employed. The effective potential calculation of Ref.~\cite{Liu:2017mrh} supported isoscalar $D^\ast\bar{D}^\ast$ and $B^\ast\bar{B}^\ast$ bound states with $J=0$, $1$ and $2$. Within the one-boson-exchange model of Ref.~\cite{Sun:2012zzd}, three isoscalar $D^\ast\bar{D}^\ast$ bound states with $J=0$, $1$, and $2$ were obtained, whereas the isovector $I(J^P)=1(0^+)$ state required a relatively large cutoff of approximately $\Lambda=3.6$ GeV. In the bottom sectors, all the systems considered were found to support bound state solutions for reasonable cutoff values. A more recent one-boson-exchange analysis~\cite{Yalikun:2025ssz}, likewise favored isoscalar $D^\ast\bar{D}^\ast$ states with $I(J^{PC})=0(0^{++})$, $0(1^{+-})$, and $0(2^++)$, with stronger binding in the corresponding bottom systems. By contrast, the complex-scaling calculation of Ref.~\cite{Lu:2025zae} found isovector $D^\ast\bar{D}^\ast$ bound states in the $I(J^{PC})=1(0^{++})$ and $1(1^{+-})$ systems. Within the quasipotential BS equation approach, Ref.~\cite{Ding:2020dio} found that all allowed channels in the $D^\ast\bar{D}^\ast$ and $B^\ast\bar{B}^\ast$ systems could support bound states.
 
Overall, most model calculations favor attraction in the isoscalar channels, particularly in the $2^{++}$ system, whereas the existence and nature of the isovector states remain strongly model dependent. The discrepancies among different predictions reflect their different treatments of short-range interactions, regulator dependence, coupled-channel effects, and model parameters.

\subsection{$D^\ast D^\ast$ and $\bar{B}^\ast \bar{B}^\ast$ systems}

\begin{figure}[htb]
\centering
\subfigure[]{
\includegraphics[width=5cm]{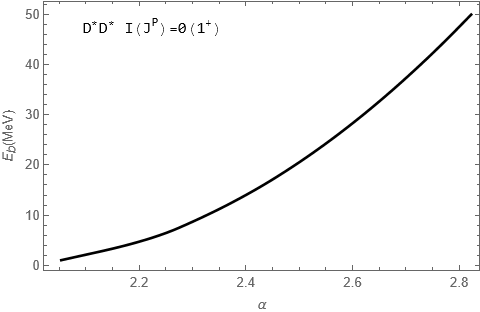}
}
\,
\subfigure[]{
\includegraphics[width=5cm]{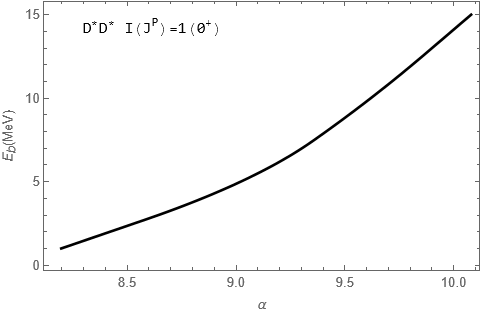}
}
\,
\subfigure[]{
\includegraphics[width=5cm]{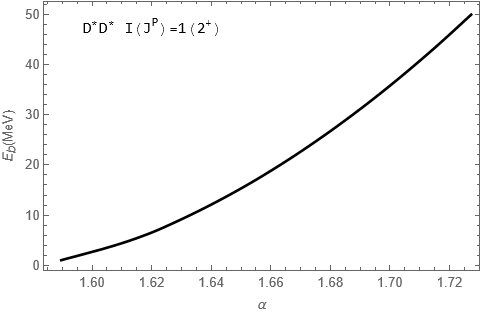}
}
\caption{Binding energy $E_b$ as a function of the cutoff parameter $\alpha$ for the possible $D^\ast D^\ast$ bound states.}
\label{Open charmed systems}
\end{figure}

\begin{figure}[htb]
\centering
\subfigure[]{
\includegraphics[width=5cm]{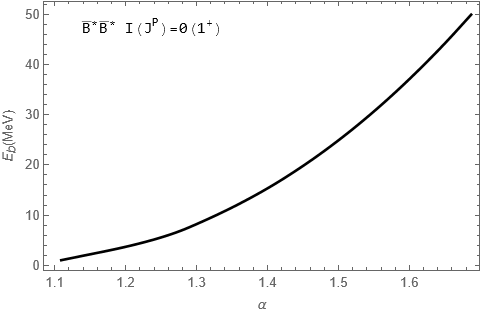}
}
\,
\subfigure[]{
\includegraphics[width=5cm]{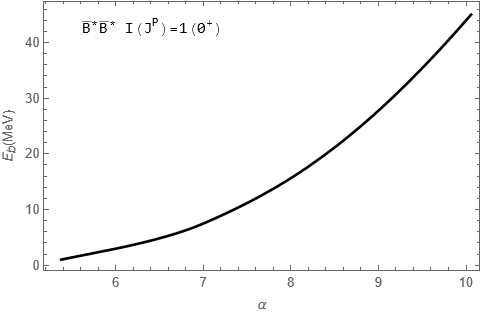}
}
\,
\subfigure[]{
\includegraphics[width=5cm]{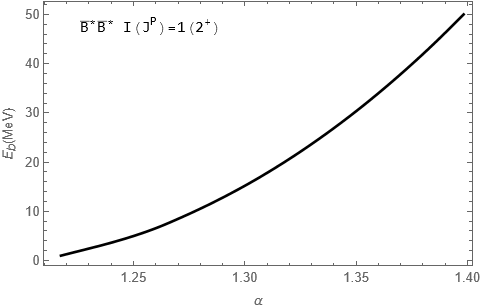}
}
\caption{Binding energy $E_b$ as a function of the cutoff parameter $\alpha$ for the possible $\bar B^\ast\bar B^\ast$ bound states.}
\label{Open bottom systems}
\end{figure}

The numerical results for the $D^\ast D^\ast$ and $\bar{B}^\ast \bar{B}^\ast$ systems are shown in Figs.~\ref{Open charmed systems} and~\ref{Open bottom systems}, respectively. Because the two constituent mesons are identical bosons, Bose symmetry restricts the allowed $S$-wave systems to have the following quantum numbers:
\[I(J^P)=0(1^+),\, 1(0^+)\,  1(2^+).\]
Within the present model, the $D^\ast D^\ast$ systems with $I(J^P)=0(1^+)$ and $1(2^+)$ support bound state solutions for moderate values of $\alpha$, whereas the $1(0^+)$ system becomes bound only for substantially large values of $\alpha$. The latter solution is therefore more sensitive to the poorly constrained short-range interaction and should be regarded as less robust. The corresponding $\bar{B}^\ast \bar{B}^\ast$ systems exhibit the same qualitative behaviour. However, for a given binding energy, they require smaller values of $\alpha$ than the $D^\ast D^\ast$ systems because their larger reduced masses reduce the kinetic-energy contribution and favor the binding. As in the hidden-heavy systems, $E_b$ increases with $\alpha$ because the larger cutoff enhances the high-momentum or short-range components of the interaction.

Several theoretical approaches have investigated these doubly heavy systems. A coupled-channel meson-exchange analysis incorporating HQSS breaking effects predicted an isoscalar $D^\ast D^\ast$ bound state with $I(J^P)=0(1^+)$~\cite{Qiu:2023uno}. An effective field theory study constrained by HQSS likewise identified the $I(J^P)=0(1^+)$ channel as the most favorable configuration in the $D^\ast D^\ast$ and $\bar{B}^\ast \bar{B}^\ast$ systems~\cite{Ren:2026wzt}. A lattice QCD study of the isovector coupled $DD$, $DD^\ast$, and $D^\ast D^\ast$ channels observed a weakly attractive $S$-wave $D^\ast D^\ast$ interaction but found no pole corresponding to a bound state or resonance in the energy region investigated~\cite{PitangaLachini:2025pxr}. This result is particularly relevant to the isovector channels and indicates that weak attraction does not necessarily generate a physical bound state. In the one-boson-exchange calculation combined with the complex-scaling method, Ref.~\cite{Lu:2025zae} obtained $D^\ast D^\ast$ bound states in the $I(J^P)=0(1^+)$ and $1(0^+)$ systems. For the bottom sector, chiral effective field theory predicted an isoscalar $\bar{B}^\ast\bar{B}^\ast$ bound state with $I(J^P)=0(1^+)$ and a binding energy of $\Delta E\simeq23.8_{-21.5}^{+16.3}$ MeV~\cite{Wang:2018atz}. A quark interchange analysis also found a bound state only in the $I=0(1^+)$ $\bar{B}^{\ast}\bar{B}^{\ast}$ channel, while the isovector $1(0^+)$ and $1(2^+)$ channels were identified as possible virtual states~\cite{Yu:2019sxx}. Within the quasipotential BS equation approach~\cite{Ding:2020dio}, the $D^\ast D^\ast$ and $\bar{B}^\ast \bar{B}^\ast$ systems with $I(J^P)=0(1^+)$ and $1(2^+)$ could be bound states.

The available studies therefore consistently identify the isoscalar $1^+$ system as the most promising doubly heavy molecular configuration. In contrast, the isovector systems are more sensitive to the treatment of short-distance dynamics and coupled-channel effects. In particular, the large value of $\alpha$ required for the $I(J^P)=1(0^+)$ solution in the present calculation suggests that this state is less natural within the adopted one-boson-exchange BS framework.

\section{Conclusions}
\label{conclusion}

In this work, we have systematically investigated possible $S$-wave molecular bound states in the $D^\ast\bar{D}^\ast$, $B^\ast\bar{B}^\ast$, $D^\ast D^\ast$, and $\bar{B}^\ast\bar{B}^\ast$ systems within the BS formalism. The BS equations were solved in the ladder and instantaneous approximations, with the interaction kernels constructed from the exchanges of $\sigma$, pseudoscalar ($\pi$ and $\eta$), and vector ($\rho$ and $\omega$) mesons. A monopole form factor was introduced at each interaction vertex to account for the finite-size effects of the constituent hadrons. The resulting integral equations were reduced to one-dimensional eigenvalue equations and solved numerically for binding energies between 1 and 50 MeV.

For the hidden-heavy systems, bound-state solutions were obtained in all the three isoscalar $D^\ast\bar{D}^\ast$ and $B^\ast\bar{B}^\ast$ channels with $J^{PC}=0^{++}$, $1^{+-}$, and $2^{++}$ when the cutoff parameters were in the reasonable range. In contrast, no isovector bound-state solution was found within the parameter range considered. These results indicate that the isoscalar interactions are generally more attractive, although the quantitative predictions depend on the treatment of short-range dynamics.

For the doubly heavy systems, Bose symmetry restricts the allowed $S$-wave channels to $I(J^P)=0(1^+)$, $1(0^+)$, and $1(2^+)$. Bound-state solutions were found in these channels for both the $D^\ast D^\ast$ and $\bar{B}^\ast\bar{B}^\ast$ systems. In particular, the $0(1^+)$ and $1(2^+)$ solutions can be generated with moderate cutoff parameters, whereas the $1(0^+)$ solution requires a substantially large cutoff and is therefore more sensitive to the poorly constrained short-range interaction. The isoscalar $1^+$ channel consequently appears to be the most promising doubly heavy molecular configuration.

For all the bound-state solutions, the binding energy increases monotonically with the cutoff parameter $\alpha$. A larger $\alpha$ corresponds to weaker suppression of high-momentum contributions by the form factor and thus enhances the short-range attraction. Moreover, the bottom systems require smaller cutoff parameters than their charmed counterparts for a given binding energy. This behavior can be attributed mainly to their larger reduced masses, which suppress the kinetic-energy contribution and facilitate binding.

Overall, our results favor isoscalar molecular states in both the hidden-heavy and doubly heavy sectors. However, solutions requiring unusually large cutoff parameters should be interpreted cautiously because of their strong dependence on short-distance dynamics which is not explicitly included in the present one-boson-exchange kernel. Further studies incorporating coupled-channel effects and additional short-range interactions, together with experimental and lattice QCD investigations, will be important for establishing the existence and properties of these molecular candidates.

\acknowledgments
This work was supported by National Natural Science Foundation of China (Project Nos. 12105149, 12405115 12475096 and 12275024) and the High-Level Scientific Research Fund of Ningbo University under Grant No. GJPY2026032.

\bibliography{ref.bib}

@article{Guo:2017jvc,
    author = "Guo, Feng-Kun and Hanhart, Christoph and Mei{\ss}ner, Ulf-G. and Wang, Qian and Zhao, Qiang and Zou, Bing-Song",
    title = "{Hadronic molecules}",
    eprint = "1705.00141",
    archivePrefix = "arXiv",
    primaryClass = "hep-ph",
    doi = "10.1103/RevModPhys.90.015004",
    journal = "Rev. Mod. Phys.",
    volume = "90",
    number = "1",
    pages = "015004",
    year = "2018",
    note = "[Erratum: Rev.Mod.Phys. 94, 029901 (2022)]"
}

@article{ParticleDataGroup:2024cfk,
    author = "Navas, S. and others",
    collaboration = "Particle Data Group",
    title = "{Review of particle physics}",
    doi = "10.1103/PhysRevD.110.030001",
    journal = "Phys. Rev. D",
    volume = "110",
    number = "3",
    pages = "030001",
    year = "2024"
}

@article{Gell-Mann:1964ewy,
    author = "Gell-Mann, Murray",
    title = "{A Schematic Model of Baryons and Mesons}",
    doi = "10.1016/S0031-9163(64)92001-3",
    journal = "Phys. Lett.",
    volume = "8",
    pages = "214--215",
    year = "1964"
}

@article{Zweig:1964ruk,
    author = "Zweig, G.",
    title = "{An SU(3) model for strong interaction symmetry and its breaking. Version 1}",
    reportNumber = "CERN-TH-401",
    doi = "10.17181/CERN-TH-401",
    journal = " ",
    volume = " ",
    pages = " ",
    month = "1",
    year = "1964"
}

@article{Belle:2003nnu,
    author = "Choi, S. K. and others",
    collaboration = "Belle",
    title = "{Observation of a narrow charmonium-like state in exclusive $B^\pm \to K^\pm \pi^+ \pi^- J/\psi$ decays}",
    eprint = "hep-ex/0309032",
    archivePrefix = "arXiv",
    doi = "10.1103/PhysRevLett.91.262001",
    journal = "Phys. Rev. Lett.",
    volume = "91",
    pages = "262001",
    year = "2003"
}

@article{Swanson:2003tb,
    author = "Swanson, Eric S.",
    title = "{Short range structure in the X(3872)}",
    eprint = "hep-ph/0311229",
    archivePrefix = "arXiv",
    reportNumber = "JLAB-THY-03-227",
    doi = "10.1016/j.physletb.2004.03.033",
    journal = "Phys. Lett. B",
    volume = "588",
    pages = "189--195",
    year = "2004"
}

@article{Maiani:2004vq,
    author = "Maiani, L. and Piccinini, F. and Polosa, A. D. and Riquer, V.",
    title = "{Diquark-antidiquarks with hidden or open charm and the nature of X(3872)}",
    eprint = "hep-ph/0412098",
    archivePrefix = "arXiv",
    reportNumber = "ROMA1-1396-2004, FNT-T-2004-20, BA-TH-502-04, CERN-PH-TH-2004-239",
    doi = "10.1103/PhysRevD.71.014028",
    journal = "Phys. Rev. D",
    volume = "71",
    pages = "014028",
    year = "2005"
}

@article{Kalashnikova:2005ui,
    author = "Kalashnikova, Yu. S.",
    title = "{Coupled-channel model for charmonium levels and an option for X(3872)}",
    eprint = "hep-ph/0506270",
    archivePrefix = "arXiv",
    doi = "10.1103/PhysRevD.72.034010",
    journal = "Phys. Rev. D",
    volume = "72",
    pages = "034010",
    year = "2005"
}

@article{Ebert:2005nc,
    author = "Ebert, D. and Faustov, R. N. and Galkin, V. O.",
    title = "{Masses of heavy tetraquarks in the relativistic quark model}",
    eprint = "hep-ph/0512230",
    archivePrefix = "arXiv",
    reportNumber = "HU-EP-05-82",
    doi = "10.1016/j.physletb.2006.01.026",
    journal = "Phys. Lett. B",
    volume = "634",
    pages = "214--219",
    year = "2006"
}

@article{Zhang:2009vs,
    author = "Zhang, Jian-Rong and Huang, Ming-Qiu",
    title = "{{Q anti-q}{anti-Q-(prime)q} molecular states}",
    eprint = "0906.0090",
    archivePrefix = "arXiv",
    primaryClass = "hep-ph",
    doi = "10.1103/PhysRevD.80.056004",
    journal = "Phys. Rev. D",
    volume = "80",
    pages = "056004",
    year = "2009"
}

@article{Chen:2009zzi,
    author = "Chen, Xiaozhao and Wang, Bin and Li, Xiaoya and Zeng, Xiaoqiang and Yu, Shaolan and Lu, Xiaofu",
    title = "{Mass of X (3872) in the relativistic quark model}",
    doi = "10.1103/PhysRevD.79.114006",
    journal = "Phys. Rev. D",
    volume = "79",
    pages = "114006",
    year = "2009"
}

@article{Ding:2009vj,
    author = "Ding, Gui-Jun and Liu, Jia-Feng and Yan, Mu-Lin",
    title = "{Dynamics of Hadronic Molecule in One-Boson Exchange Approach and Possible Heavy Flavor Molecules}",
    eprint = "0901.0426",
    archivePrefix = "arXiv",
    primaryClass = "hep-ph",
    doi = "10.1103/PhysRevD.79.054005",
    journal = "Phys. Rev. D",
    volume = "79",
    pages = "054005",
    year = "2009"
}

@article{Liu:2008qb,
    author = "Liu, Yan-Rui and Zhang, Zong-Ye",
    title = "{X(3872) and the bound state problem of $D^0 \bar{D}$* 0 (anti-D0 D*0) in a chiral quark model}",
    eprint = "0805.1616",
    archivePrefix = "arXiv",
    primaryClass = "hep-ph",
    doi = "10.1103/PhysRevC.79.035206",
    journal = "Phys. Rev. C",
    volume = "79",
    pages = "035206",
    year = "2009"
}

@article{Carlucci:2007um,
    author = "Carlucci, M. V. and Giannuzzi, Floriana and Nardulli, G. and Pellicoro, M. and Stramaglia, S.",
    title = "{AdS-QCD quark-antiquark potential, meson spectrum and tetraquarks}",
    eprint = "0711.2014",
    archivePrefix = "arXiv",
    primaryClass = "hep-ph",
    reportNumber = "BARI-TH-584-07",
    doi = "10.1140/epjc/s10052-008-0687-2",
    journal = "Eur. Phys. J. C",
    volume = "57",
    pages = "569--578",
    year = "2008"
}

@article{Vijande:2007fc,
    author = "Vijande, J. and Weissman, E. and Barnea, N. and Valcarce, A.",
    title = "{Do $c \bar{c} n \bar{n}$ bound states exist?}",
    eprint = "0708.3285",
    archivePrefix = "arXiv",
    primaryClass = "hep-ph",
    doi = "10.1103/PhysRevD.76.094022",
    journal = "Phys. Rev. D",
    volume = "76",
    pages = "094022",
    year = "2007"
}

@article{Maiani:2007vr,
    author = "Maiani, L. and Polosa, A. D. and Riquer, V.",
    title = "{Indications of a Four-Quark Structure for the X(3872) and X(3876) Particles from Recent Belle and BABAR Data}",
    eprint = "0707.3354",
    archivePrefix = "arXiv",
    primaryClass = "hep-ph",
    doi = "10.1103/PhysRevLett.99.182003",
    journal = "Phys. Rev. Lett.",
    volume = "99",
    pages = "182003",
    year = "2007"
}

@article{Nieves:2012tt,
    author = "Nieves, J. and Valderrama, M. Pavon",
    title = "{The Heavy Quark Spin Symmetry Partners of the X(3872)}",
    eprint = "1204.2790",
    archivePrefix = "arXiv",
    primaryClass = "hep-ph",
    doi = "10.1103/PhysRevD.86.056004",
    journal = "Phys. Rev. D",
    volume = "86",
    pages = "056004",
    year = "2012"
}

@article{Dubnicka:2010kz,
    author = "Dubnicka, Stanislav and Dubnickova, Anna Z. and Ivanov, Mikhail A. and Korner, Juergen G.",
    title = "{Quark model description of the tetraquark state X(3872) in a relativistic constituent quark model with infrared confinement}",
    eprint = "1004.1291",
    archivePrefix = "arXiv",
    primaryClass = "hep-ph",
    doi = "10.1103/PhysRevD.81.114007",
    journal = "Phys. Rev. D",
    volume = "81",
    pages = "114007",
    year = "2010"
}

@article{Lee:2009hy,
    author = "Lee, Ian Woo and Faessler, Amand and Gutsche, Thomas and Lyubovitskij, Valery E.",
    title = "{X(3872) as a molecular DD* state in a potential model}",
    eprint = "0910.1009",
    archivePrefix = "arXiv",
    primaryClass = "hep-ph",
    doi = "10.1103/PhysRevD.80.094005",
    journal = "Phys. Rev. D",
    volume = "80",
    pages = "094005",
    year = "2009"
}

@article{Guo:2014taa,
    author = "Guo, Feng-Kun and Hanhart, C. and Kalashnikova, Yu. S. and Mei{\ss}ner, Ulf-G. and Nefediev, A. V.",
    title = "{What can radiative decays of the X(3872) teach us about its nature?}",
    eprint = "1410.6712",
    archivePrefix = "arXiv",
    primaryClass = "hep-ph",
    doi = "10.1016/j.physletb.2015.02.013",
    journal = "Phys. Lett. B",
    volume = "742",
    pages = "394--398",
    year = "2015"
}

@article{He:2014nya,
    author = "He, Jun",
    title = "{Study of the $B\bar{B}^*/D\bar{D}^*$ bound states in a Bethe-Salpeter approach}",
    eprint = "1409.8506",
    archivePrefix = "arXiv",
    primaryClass = "hep-ph",
    doi = "10.1103/PhysRevD.90.076008",
    journal = "Phys. Rev. D",
    volume = "90",
    number = "7",
    pages = "076008",
    year = "2014"
}

@article{Zhao:2014gqa,
    author = "Zhao, Lu and Ma, Li and Zhu, Shi-Lin",
    title = "{Spin-orbit force, recoil corrections, and possible $B \bar{B}^{*}$ and $D \bar{D}^{*}$  molecular states}",
    eprint = "1403.4043",
    archivePrefix = "arXiv",
    primaryClass = "hep-ph",
    doi = "10.1103/PhysRevD.89.094026",
    journal = "Phys. Rev. D",
    volume = "89",
    number = "9",
    pages = "094026",
    year = "2014"
}

@article{Wang:2013vex,
    author = "Wang, Zhi-Gang and Huang, Tao",
    title = "{Analysis of the $X(3872)$, $Z_c(3900)$ and $Z_c(3885)$ as axial-vector tetraquark states with QCD sum rules}",
    eprint = "1310.2422",
    archivePrefix = "arXiv",
    primaryClass = "hep-ph",
    doi = "10.1103/PhysRevD.89.054019",
    journal = "Phys. Rev. D",
    volume = "89",
    number = "5",
    pages = "054019",
    year = "2014"
}

@article{Guo:2013sya,
    author = "Guo, Feng-Kun and Hidalgo-Duque, Carlos and Nieves, Juan and Valderrama, Manuel Pavon",
    title = "{Consequences of Heavy Quark Symmetries for Hadronic Molecules}",
    eprint = "1303.6608",
    archivePrefix = "arXiv",
    primaryClass = "hep-ph",
    doi = "10.1103/PhysRevD.88.054007",
    journal = "Phys. Rev. D",
    volume = "88",
    pages = "054007",
    year = "2013"
}

@article{Hidalgo-Duque:2012rqv,
    author = "Hidalgo-Duque, C. and Nieves, J. and Valderrama, M. Pavon",
    title = "{Light flavor and heavy quark spin symmetry in heavy meson molecules}",
    eprint = "1210.5431",
    archivePrefix = "arXiv",
    primaryClass = "hep-ph",
    doi = "10.1103/PhysRevD.87.076006",
    journal = "Phys. Rev. D",
    volume = "87",
    number = "7",
    pages = "076006",
    year = "2013"
}

@article{Baru:2016iwj,
    author = "Baru, V. and Epelbaum, E. and Filin, A. A. and Hanhart, C. and Mei{\ss}ner, Ulf-G. and Nefediev, A. V.",
    title = "{Heavy-quark spin symmetry partners of the X (3872) revisited}",
    eprint = "1605.09649",
    archivePrefix = "arXiv",
    primaryClass = "hep-ph",
    doi = "10.1016/j.physletb.2016.10.008",
    journal = "Phys. Lett. B",
    volume = "763",
    pages = "20--28",
    year = "2016"
}

@article{Padmanath:2015era,
    author = "Padmanath, M. and Lang, C. B. and Prelovsek, Sasa",
    title = "{X(3872) and Y(4140) using diquark-antidiquark operators with lattice QCD}",
    eprint = "1503.03257",
    archivePrefix = "arXiv",
    primaryClass = "hep-lat",
    reportNumber = "JLAB-THY-15-2017",
    doi = "10.1103/PhysRevD.92.034501",
    journal = "Phys. Rev. D",
    volume = "92",
    number = "3",
    pages = "034501",
    year = "2015"
}

@article{Wang:2017dcq,
    author = "Wang, Zhen-Yang and Qi, Jing-Juan and Guo, Xin-Heng and Wang, Chao",
    title = "{X(3872) as a molecular $D\bar{D}^*$ state in the Bethe-Salpeter equation approach}",
    eprint = "1710.07424",
    archivePrefix = "arXiv",
    primaryClass = "hep-ph",
    doi = "10.1103/PhysRevD.97.016015",
    journal = "Phys. Rev. D",
    volume = "97",
    number = "1",
    pages = "016015",
    year = "2018"
}

@article{Anwar:2018sol,
    author = "Anwar, Muhammad Naeem and Ferretti, Jacopo and Santopinto, Elena",
    title = "{Spectroscopy of the hidden-charm $[qc][\bar q \bar c]$ and $[sc][\bar s \bar c]$ tetraquarks in the relativized diquark model}",
    eprint = "1805.06276",
    archivePrefix = "arXiv",
    primaryClass = "hep-ph",
    doi = "10.1103/PhysRevD.98.094015",
    journal = "Phys. Rev. D",
    volume = "98",
    number = "9",
    pages = "094015",
    year = "2018"
}

@article{Wallbott:2019dng,
    author = "Wallbott, Paul C. and Eichmann, Gernot and Fischer, Christian S.",
    title = "{$X(3872)$ as a four-quark state in a Dyson-Schwinger/Bethe-Salpeter approach}",
    eprint = "1905.02615",
    archivePrefix = "arXiv",
    primaryClass = "hep-ph",
    doi = "10.1103/PhysRevD.100.014033",
    journal = "Phys. Rev. D",
    volume = "100",
    number = "1",
    pages = "014033",
    year = "2019"
}

@article{Zhao:2021cvg,
    author = "Zhao, Meng-Jie and Wang, Zhen-Yang and Wang, Chao and Guo, Xin-Heng",
    title = "{Investigation of the possible DD{\textasciimacron}*/BB{\textasciimacron}* and DD*/B{\textasciimacron}B{\textasciimacron}* bound states}",
    eprint = "2112.12633",
    archivePrefix = "arXiv",
    primaryClass = "hep-ph",
    doi = "10.1103/PhysRevD.105.096016",
    journal = "Phys. Rev. D",
    volume = "105",
    number = "9",
    pages = "096016",
    year = "2022"
}

@article{Lu:2025zae,
    author = "Lu, Jia-Liang and Song, Mao and Wang, Peng and Guo, Jian-You and Li, Gang and Luo, Xuan",
    title = "{The bound and resonant states of $D^{(*)}D^{(*)}$ and $D^{(*)}{\bar{D}}^{(*)}$ with the complex scaling method}",
    eprint = "2503.05131",
    archivePrefix = "arXiv",
    primaryClass = "hep-ph",
    doi = "10.1140/epjc/s10052-025-14649-1",
    journal = "Eur. Phys. J. C",
    volume = "85",
    number = "8",
    pages = "920",
    year = "2025"
}

@article{Li:2024pfg,
    author = "Li, Haozheng and Shi, Chunjiang and Chen, Ying and Gong, Ming and Liang, Juzheng and Liu, Zhaofeng and Sun, Wei",
    title = "{$X(3872)$ Relevant $D\bar{D}^\ast$ Scattering in $N_f=2$ Lattice QCD}",
    eprint = "2402.14541",
    archivePrefix = "arXiv",
    primaryClass = "hep-lat",
    doi = " ",
    journal = " ",
    volume = " ",
    number = " ",
    pages = " ",
    month = "2",
    year = "2024"
}

@article{Belle:2005lik,
    author = "Abe, Kazuo and others",
    collaboration = "Belle",
    title = "{Observation of a new charmonium state in double charmonium production in $e^+ e^-$ annihilation at $\sqrt{s}$ $\approx$ 10.6-GeV}",
    eprint = "hep-ex/0507019",
    archivePrefix = "arXiv",
    reportNumber = "BELLE-CONF-0517, LP2005-152, EPS05-489",
    doi = "10.1103/PhysRevLett.98.082001",
    journal = "Phys. Rev. Lett.",
    volume = "98",
    pages = "082001",
    year = "2007"
}

@article{BESIII:2013ris,
    author = "Ablikim, M. and others",
    collaboration = "BESIII",
    title = "{Observation of a Charged Charmoniumlike Structure in $e^+e^- \to \pi^+\pi^- J/\psi$ at $\sqrt{s}$ =4.26 GeV}",
    eprint = "1303.5949",
    archivePrefix = "arXiv",
    primaryClass = "hep-ex",
    doi = "10.1103/PhysRevLett.110.252001",
    journal = "Phys. Rev. Lett.",
    volume = "110",
    pages = "252001",
    year = "2013"
}

@article{Belle:2007woe,
    author = "Pakhlov, P. and others",
    collaboration = "Belle",
    title = "{Production of New Charmoniumlike States in $e^+ e^- \to  J/\psi D^{(\ast)} \bar{D}^{(\ast)}$ at $\sqrt{S} \approx$ 10.6 GeV}",
    eprint = "0708.3812",
    archivePrefix = "arXiv",
    primaryClass = "hep-ex",
    doi = "10.1103/PhysRevLett.100.202001",
    journal = "Phys. Rev. Lett.",
    volume = "100",
    pages = "202001",
    year = "2008"
}

@article{BESIII:2013ouc,
    author = "Ablikim, M. and others",
    collaboration = "BESIII",
    title = "{Observation of a Charged Charmoniumlike Structure $Z_c$(4020) and Search for the $Z_c$(3900) in $e^+e^- \to \pi^+\pi^- h_c$}",
    eprint = "1309.1896",
    archivePrefix = "arXiv",
    primaryClass = "hep-ex",
    doi = "10.1103/PhysRevLett.111.242001",
    journal = "Phys. Rev. Lett.",
    volume = "111",
    number = "24",
    pages = "242001",
    year = "2013"
}

@article{LHCb:2021vvq,
    author = "Aaij, Roel and others",
    collaboration = "LHCb",
    title = "{Observation of an exotic narrow doubly charmed tetraquark}",
    eprint = "2109.01038",
    archivePrefix = "arXiv",
    primaryClass = "hep-ex",
    reportNumber = "CERN-EP-2021-165, LHCb-PAPER-2021-031",
    doi = "10.1038/s41567-022-01614-y",
    journal = "Nature Phys.",
    volume = "18",
    number = "7",
    pages = "751--754",
    year = "2022"
}

@article{LHCb:2021auc,
    author = "Aaij, Roel and others",
    collaboration = "LHCb",
    title = "{Study of the doubly charmed tetraquark $T_{cc}^{+}$}",
    eprint = "2109.01056",
    archivePrefix = "arXiv",
    primaryClass = "hep-ex",
    reportNumber = "CERN-EP-2021-169, LHCb-PAPER-2021-032",
    doi = "10.1038/s41467-022-30206-w",
    journal = "Nature Commun.",
    volume = "13",
    number = "1",
    pages = "3351",
    year = "2022"
}

@article{Belle:2011aa,
    author = "Bondar, A. and others",
    collaboration = "Belle",
    title = "{Observation of two charged bottomonium-like resonances in Y(5S) decays}",
    eprint = "1110.2251",
    archivePrefix = "arXiv",
    primaryClass = "hep-ex",
    doi = "10.1103/PhysRevLett.108.122001",
    journal = "Phys. Rev. Lett.",
    volume = "108",
    pages = "122001",
    year = "2012"
}

@article{Dong:2020hxe,
    author = "Dong, Xiang-Kun and Guo, Feng-Kun and Zou, Bing-Song",
    title = "{Explaining the Many Threshold Structures in the Heavy-Quark Hadron Spectrum}",
    eprint = "2011.14517",
    archivePrefix = "arXiv",
    primaryClass = "hep-ph",
    doi = "10.1103/PhysRevLett.126.152001",
    journal = "Phys. Rev. Lett.",
    volume = "126",
    number = "15",
    pages = "152001",
    year = "2021"
}

@article{Chen:2016qju,
    author = "Chen, Hua-Xing and Chen, Wei and Liu, Xiang and Zhu, Shi-Lin",
    title = "{The hidden-charm pentaquark and tetraquark states}",
    eprint = "1601.02092",
    archivePrefix = "arXiv",
    primaryClass = "hep-ph",
    doi = "10.1016/j.physrep.2016.05.004",
    journal = "Phys. Rept.",
    volume = "639",
    pages = "1--121",
    year = "2016"
}

@article{Brambilla:2019esw,
    author = "Brambilla, Nora and Eidelman, Simon and Hanhart, Christoph and Nefediev, Alexey and Shen, Cheng-Ping and Thomas, Christopher E. and Vairo, Antonio and Yuan, Chang-Zheng",
    title = "{The $XYZ$ states: experimental and theoretical status and perspectives}",
    eprint = "1907.07583",
    archivePrefix = "arXiv",
    primaryClass = "hep-ex",
    reportNumber = "TUM-EFT 125/19",
    doi = "10.1016/j.physrep.2020.05.001",
    journal = "Phys. Rept.",
    volume = "873",
    pages = "1--154",
    year = "2020"
}

@article{Liu:2024uxn,
    author = "Liu, Ming-Zhu and Pan, Ya-Wen and Liu, Zhi-Wei and Wu, Tian-Wei and Lu, Jun-Xu and Geng, Li-Sheng",
    title = "{Three ways to decipher the nature of exotic hadrons: Multiplets, three-body hadronic molecules, and correlation functions}",
    eprint = "2404.06399",
    archivePrefix = "arXiv",
    primaryClass = "hep-ph",
    doi = "10.1016/j.physrep.2024.12.001",
    journal = "Phys. Rept.",
    volume = "1108",
    pages = "1--108",
    year = "2025"
}

@article{Ding:2020dio,
    author = "Ding, Zuo-Ming and Jiang, Han-Yu and He, Jun",
    title = "{Molecular states from $D^{(*)}\bar{D}^{(*)}/B^{(*)}\bar{B}^{(*)}$ and $D^{(*)}D^{(*)}/\bar{B}^{(*)}\bar{B}^{(*)}$ interactions}",
    eprint = "2011.04980",
    archivePrefix = "arXiv",
    primaryClass = "hep-ph",
    doi = "10.1140/epjc/s10052-020-08754-6",
    journal = "Eur. Phys. J. C",
    volume = "80",
    number = "12",
    pages = "1179",
    year = "2020"
}

@article{Dong:2021juy,
    author = "Dong, Xiang-Kun and Guo, Feng-Kun and Zou, Bing-Song",
    title = "{A survey of heavy-antiheavy hadronic molecules}",
    eprint = "2101.01021",
    archivePrefix = "arXiv",
    primaryClass = "hep-ph",
    doi = "10.13725/j.cnki.pip.2021.02.001",
    journal = "Progr. Phys.",
    volume = "41",
    pages = "65--93",
    year = "2021"
}

@article{Dong:2021bvy,
    author = "Dong, Xiang-Kun and Guo, Feng-Kun and Zou, Bing-Song",
    title = "{A survey of heavy{\textendash}heavy hadronic molecules}",
    eprint = "2108.02673",
    archivePrefix = "arXiv",
    primaryClass = "hep-ph",
    doi = "10.1088/1572-9494/ac27a2",
    journal = "Commun. Theor. Phys.",
    volume = "73",
    number = "12",
    pages = "125201",
    year = "2021"
}

@article{Ke:2021rxd,
    author = "Ke, Hong-Wei and Liu, Xiao-Hai and Li, Xue-Qian",
    title = "{Possible molecular states of $D^{(*)}D^{(*)}$ and $B^{(*)}B^{(*)}$ within the Bethe{\textendash}Salpeter framework}",
    eprint = "2112.14142",
    archivePrefix = "arXiv",
    primaryClass = "hep-ph",
    doi = "10.1140/epjc/s10052-022-10092-8",
    journal = "Eur. Phys. J. C",
    volume = "82",
    number = "2",
    pages = "144",
    year = "2022"
}

@article{Yalikun:2025ssz,
    author = "Yalikun, Nijiati and Dong, Xiang-Kun and Mei{\ss}ner, Ulf-G.",
    title = "{Role of the short-range dynamics in the formation of D(*)D{\textasciimacron}(*) and B(*)B{\textasciimacron}(*) hadronic molecules}",
    eprint = "2503.01322",
    archivePrefix = "arXiv",
    primaryClass = "hep-ph",
    doi = "10.1103/PhysRevD.111.094036",
    journal = "Phys. Rev. D",
    volume = "111",
    number = "9",
    pages = "094036",
    year = "2025"
}

@article{Abreu:2025mhl,
    author = "Abreu, Luciano M.",
    title = "{Production of the X(3872) state via the B0{\textrightarrow}K*0X(3872) decay}",
    eprint = "2508.21223",
    archivePrefix = "arXiv",
    primaryClass = "hep-ph",
    doi = "10.1103/76ll-n7yw",
    journal = "Phys. Rev. D",
    volume = "112",
    number = "9",
    pages = "096002",
    year = "2025"
}

@article{Cai:2025inq,
    author = "Cai, Hao-Dong and Jia, Zhao-Sai and Li, Gang and Liu, Shi-Dong",
    title = "{Hidden charmed decays of X(3872) within the DD{\textasciimacron}* molecular framework}",
    eprint = "2503.20183",
    archivePrefix = "arXiv",
    primaryClass = "hep-ph",
    doi = "10.1103/gn25-fc9q",
    journal = "Phys. Rev. D",
    volume = "111",
    number = "11",
    pages = "114024",
    year = "2025"
}

@article{Dias:2024zfh,
    author = "Dias, Jorgivan Morais and Ji, Teng and Dong, Xiang-Kun and Guo, Feng-Kun and Hanhart, Christoph and Mei{\ss}ner, Ulf-G. and Zhang, Yu and Zhang, Zhen-Hua",
    title = "{Dispersive analysis of the isospin breaking in the X(3872){\textrightarrow}J/{\ensuremath{\psi}}{\ensuremath{\pi}}+{\ensuremath{\pi}}- and X(3872){\textrightarrow}J/{\ensuremath{\psi}}{\ensuremath{\pi}}+{\ensuremath{\pi}}0{\ensuremath{\pi}}- decays}",
    eprint = "2409.13245",
    archivePrefix = "arXiv",
    primaryClass = "hep-ph",
    doi = "10.1103/PhysRevD.111.014031",
    journal = "Phys. Rev. D",
    volume = "111",
    number = "1",
    pages = "014031",
    year = "2025"
}

@article{Liu:2024ziu,
    author = "Liu, Ming-Zhu and Ling, Xi-Zhe and Geng, Li-Sheng",
    title = "{Productions of X(3872)/Zc(3900) and X2(4013)/Zc(4020) in Y(4220) and Y(4360) decays}",
    eprint = "2404.07681",
    archivePrefix = "arXiv",
    primaryClass = "hep-ph",
    doi = "10.1103/PhysRevD.110.054035",
    journal = "Phys. Rev. D",
    volume = "110",
    number = "5",
    pages = "054035",
    year = "2024"
}

@article{Deng:2024pep,
    author = "Deng, Jin-Cheng and Wang, Bo",
    title = "{Strong decays of the isovector-scalar D*D{\textasciimacron}* hadronic molecule}",
    eprint = "2406.07000",
    archivePrefix = "arXiv",
    primaryClass = "hep-ph",
    doi = "10.1103/PhysRevD.110.054014",
    journal = "Phys. Rev. D",
    volume = "110",
    number = "5",
    pages = "054014",
    year = "2024"
}

@article{Wu:2023rrp,
    author = "Wu, Qi and Liu, Ming-Zhu and Geng, Li-Sheng",
    title = "{Productions of X(3872), $Z_c(3900)$, $X_2(4013)$, and $Z_c(4020)$ in $B_{(s)}$ decays offer strong clues on their molecular nature}",
    eprint = "2304.05269",
    archivePrefix = "arXiv",
    primaryClass = "hep-ph",
    doi = "10.1140/epjc/s10052-024-12501-6",
    journal = "Eur. Phys. J. C",
    volume = "84",
    number = "2",
    pages = "147",
    year = "2024"
}

@article{Wang:2022aiu,
    author = "Wang, Xiao-Yun and Li, Gang and An, Chun-Sheng and Xie, Ju-Jun",
    title = "{Radiative decays of the neutral Zc(3900) and Zc(4020)}",
    eprint = "2210.06783",
    archivePrefix = "arXiv",
    primaryClass = "hep-ph",
    doi = "10.1103/PhysRevD.106.074026",
    journal = "Phys. Rev. D",
    volume = "106",
    number = "7",
    pages = "074026",
    year = "2022"
}

@article{Zhang:2025rlg,
    author = "Zhang, Yi and Liu, Ming-Zhu and Geng, Li-Sheng",
    title = "{Productions of Tcc and its SU(3)-flavor symmetry and heavy quark spin symmetry partners in Bc decays}",
    eprint = "2510.11078",
    archivePrefix = "arXiv",
    primaryClass = "hep-ph",
    doi = "10.1103/891b-jzpy",
    journal = "Phys. Rev. D",
    volume = "113",
    number = "3",
    pages = "034029",
    year = "2026"
}

@article{Qi:2024dqz,
    author = "Qi, Jing-Juan and Zhang, Zhen-Hua and Guo, Xin-Heng and Wang, Zhen-Yang",
    title = "{Possible bound states of heavy baryonium and heavy dibaryon systems}",
    eprint = "2409.03315",
    archivePrefix = "arXiv",
    primaryClass = "hep-ph",
    doi = "10.1103/PhysRevD.110.094050",
    journal = "Phys. Rev. D",
    volume = "110",
    number = "9",
    pages = "094050",
    year = "2024"
}

@article{Wong:2003xk,
    author = "Wong, Cheuk-Yin",
    title = "{Molecular states of heavy quark mesons}",
    eprint = "hep-ph/0311088",
    archivePrefix = "arXiv",
    doi = "10.1103/PhysRevC.69.055202",
    journal = "Phys. Rev. C",
    volume = "69",
    pages = "055202",
    year = "2004"
}

@article{Liu:2017mrh,
    author = "Liu, Ming-Zhu and Jia, Duo-Jie and Chen, Dian-Yong",
    title = "{Possible hadronic molecular states composed of $S$-wave heavy-light mesons}",
    eprint = "1702.04440",
    archivePrefix = "arXiv",
    primaryClass = "hep-ph",
    doi = "10.1088/1674-1137/41/5/053105",
    journal = "Chin. Phys. C",
    volume = "41",
    number = "5",
    pages = "053105",
    year = "2017"
}

@article{Liu:2008mi,
    author = "Liu, Yan-Rui and Zhang, Zong-Ye",
    title = "{The Bound state problem of S-wave heavy quark meson-aitimeson systems}",
    eprint = "0810.1598",
    archivePrefix = "arXiv",
    primaryClass = "hep-ph",
    doi = "10.1103/PhysRevC.80.015208",
    journal = "Phys. Rev. C",
    volume = "80",
    pages = "015208",
    year = "2009"
}

@article{Ding:2009zq,
    author = "Ding, Gui-Jun",
    title = "{Bound States of the Heavy Flavor Vector Mesons and Y(4008) and Z+(1)(4050)}",
    eprint = "0905.1188",
    archivePrefix = "arXiv",
    primaryClass = "hep-ph",
    doi = "10.1103/PhysRevD.80.034005",
    journal = "Phys. Rev. D",
    volume = "80",
    pages = "034005",
    year = "2009"
}

@article{Sun:2012zzd,
    author = "Sun, Zhi-Feng and Luo, Zhi-Gang and He, Jun and Liu, Xiang and Zhu, Shi-Lin",
    title = "{A note on the B* anti-B, B* anti-B*, D* anti-D, D* anti-D*, molecular states}",
    doi = "10.1088/1674-1137/36/3/002",
    journal = "Chin. Phys. C",
    volume = "36",
    pages = "194--204",
    year = "2012"
}

@article{Wang:2018atz,
    author = "Wang, Bo and Liu, Zhan-Wei and Liu, Xiang",
    title = "{$\bar{B}^{(\ast)} \bar{B}^{(\ast)}$ interactions in chiral effective field theory}",
    eprint = "1812.04457",
    archivePrefix = "arXiv",
    primaryClass = "hep-ph",
    doi = "10.1103/PhysRevD.99.036007",
    journal = "Phys. Rev. D",
    volume = "99",
    number = "3",
    pages = "036007",
    year = "2019"
}

@article{Yu:2019sxx,
    author = "Yu, Meng-Ting and Zhou, Zhi-Yong and Chen, Dian-Yong and Xiao, Zhiguang",
    title = "{Possible molecular states in $B^{(*)}B^{(*)}$ scatterings}",
    eprint = "1912.07348",
    archivePrefix = "arXiv",
    primaryClass = "hep-ph",
    doi = "10.1103/PhysRevD.101.074027",
    journal = "Phys. Rev. D",
    volume = "101",
    number = "7",
    pages = "074027",
    year = "2020"
}

@article{Liu:2009qhy,
    author = "Liu, Xiang and Luo, Zhi-Gang and Liu, Yan-Rui and Zhu, Shi-Lin",
    title = "{X(3872) and Other Possible Heavy Molecular States}",
    eprint = "0808.0073",
    archivePrefix = "arXiv",
    primaryClass = "hep-ph",
    doi = "10.1140/epjc/s10052-009-1020-4",
    journal = "Eur. Phys. J. C",
    volume = "61",
    pages = "411--428",
    year = "2009"
}

@article{Qiu:2023uno,
    author = "Qiu, Lin and Gong, Chang and Zhao, Qiang",
    title = "{Coupled-channel description of charmed heavy hadronic molecules within the meson-exchange model and its implication}",
    eprint = "2311.10067",
    archivePrefix = "arXiv",
    primaryClass = "hep-ph",
    doi = "10.1103/PhysRevD.109.076016",
    journal = "Phys. Rev. D",
    volume = "109",
    number = "7",
    pages = "076016",
    year = "2024"
}

@article{Ren:2026wzt,
    author = "Ren, Yu-Shan and Wang, Guang-Juan and Yang, Zhi and Wu, Jia-Jun",
    title = "{Systematic Study of Coupled-Channel Dynamics in Doubly Heavy Hadronic Molecules}",
    eprint = "2606.01177",
    archivePrefix = "arXiv",
    primaryClass = "hep-ph",
    doi = " ",
    journal = " ",
    volume = " ",
    number = " ",
    pages = "",
    month = "5",
    year = "2026"
}

@article{PitangaLachini:2025pxr,
    author = "Pitanga Lachini, Nelson and Thomas, Christopher E. and Wilson, David J.",
    collaboration = "Hadron Spectrum",
    title = "{Coupled-channel scattering of DD, DD*, and D*D* in isospin-1 from lattice QCD}",
    eprint = "2505.01363",
    archivePrefix = "arXiv",
    primaryClass = "hep-lat",
    doi = "10.1103/xz5r-p533",
    journal = "Phys. Rev. D",
    volume = "112",
    number = "11",
    pages = "114511",
    year = "2025"
}

\end{document}